\documentclass[aps,pr
,twocolumn,english,showpacs,superscriptaddress,amssymb,amsfonts]{revtex4}
\usepackage[T1]{fontenc}
\usepackage[latin9]{inputenc}
\usepackage{amsmath}
\usepackage{dsfont}
\usepackage{amstext}

\usepackage{tocvsec2}

\usepackage{amssymb}
\usepackage{amsbsy}
\usepackage{amsthm}
\usepackage{epsfig}
\usepackage{framed}
\usepackage{graphicx}
\usepackage{bbm}
\usepackage{hyperref}
\usepackage{color}
\usepackage{multirow}

\newcommand{\Ref}[1]{Ref.~\onlinecite{#1}}

\newcommand{\bst}{{\mathcal{T}}}

\newcommand{\bsi}{{\mathcal{I}}}

\newcommand{\ie}{{\emph{i.e.~}}}
\makeatletter

\newcommand{\Rmnum}[1]{\expandafter\@slowromancap\romannumeral #1@}
\makeatother
\newcommand{\imth}{\hspace{1pt}\mathrm{i}\hspace{1pt}}

\newcommand{\eg}{{\emph{e.g.~}}}

\newcommand{\mbz}{{\mathbb{Z}}}
\newcommand{\bea}{\begin{eqnarray}}
\newcommand{\eea}{\end{eqnarray}}
\newcommand{\bpm}{\begin{pmatrix}}
\newcommand{\epm}{\end{pmatrix}}
\newcommand{\bal}{\begin{aligned}}
\newcommand{\eal}{\end{aligned}}

\newcommand{\expval}[1]{\langle{#1}\rangle}

\newcommand{\gpc}[3]{{\mathcal{H}^{#1}\big({#2},{#3}\big)}}

\makeatother

\usepackage{babel}
\begin{document}
\title{Intrinsically interacting topological crystalline insulators and superconductors}

\author{Alex Rasmussen}
\author{Yuan-Ming Lu}
\affiliation{Department of Physics, The Ohio State University, Columbus, OH 43210, USA}

\begin{abstract}
Motivated by recent progress in crystalline symmetry protected topological (SPT) phases of interacting bosons, we study topological crystalline insulators/superconductors (TCIs) of strongly interacting fermions. We construct a class of intrinsically interacting fermionic TCIs, and show that they are beyond both free-fermion TCIs and bosonic crystalline SPT phases. We also show how these phases can be characterized by symmetry protected gapless fermion modes on the corners/hinges of an open system. 
\end{abstract}

%\hpacs{}

\maketitle
%\setcounter{tocdepth}{2}
%\begin{widetext}
%\tableofcontents
%\end{widetext}

%\maxsecnumdepth{subsection}

%{\small \setcounter{tocdepth}{2} \tableofcontents}

%\twocolumn

\section{Introduction}

The discovery of topological insulators (TIs)\cite{Hasan2010,Hasan2011,Qi2011c} and their counterparts in interacting bosons, symmetry-protected topological states (SPTs)\cite{Chen2013,Senthil2015}, revealed a large class of topological phases with symmetry protected topological boundary states despite a gapped trivial bulk. While SPTs are well understood and classified by K-theory\cite{Schnyder2008,Kitaev2009,Chiu2016} for free fermions and by group cohomology\cite{Chen2013} and cobordism\cite{Kapustin2014} for interacting bosons, less is known about a full classification of interacting fermion SPTs\cite{Gu2014,Kapustin2015,Kapustin2017,Wang2018d,Cheng2018a,Lan2018}. In a system of interacting fermions, in addition to free-fermion SPTs, it has been found that an even number of fermions can also form a bosonic bound state which in turn forms a bosonic SPT phase\cite{Lu2012a,Wang2015}. This raises a natural question: are there any interacting fermionic SPTs, which cannot be realized by stacking free fermion SPTs and bosonic SPTs? Recently it has been argued based on braiding statistics that such intrinsically interacting SPTs do exist\cite{Wang2017,Cheng2018} in two and three dimensions, although it is not clear how to realize them in concrete lattice models of interacting fermions. 

In this work, we explicitly construct a class of intrinsically interacting SPTs of fermions, protected by both global (``onsite'') and crystalline symmetries. These phases are coined topological crystallline insulators (TCIs) and superconductors\cite{Ando2015} in the context of TIs, hence we will call them ``intrinsically interacting TCIs'' (for both insulators and superconductors) throughout this work. This work is inspired by recent progress in classifying bosonic SPT phases with both onsite and crystalline symmetries\cite{Thorngren2018,Jiang2017,Song2017a,Huang2017b,Lu2017c,Rasmussen2018,Shiozaki2018,Else2018,Song2018a}, which points to a dimensional reduction scheme to construct interacting TCIs. In particular, inspired by recent progress on higher-order SPT phases\cite{Parameswaran2017,Benalcazar2017,Benalcazar2017a,Song2017,Langbehn2017}, we show these intrinsically interacting TCIs are characterized by robust fermion modes on the corners/hinges of an open system, which serves as a topological invariant differentiating these interacting TCIs from free-fermion states and bosonic SPTs. 

The paper is organized as follows. In Sec. \ref{sec:general}, we first lay out the general strategy behind the decorated domain wall construction for intrinsically interacting TCIs. Next we explicitly construct 3 examples in two (2d) and three (3d) dimensions in Sec. \ref{sec:k=1,d=2}-\ref{sec:k=2,d=3}, and establish that they are neither free fermion nor bosonic SPTs. Finally we discuss limitations of our current construction and future directions in Sec. \ref{sec:discussion}. 

\section{General strategy}\label{sec:general}

Before explicitly constructing the interacting TCIs, we outline the general strategy for our construction, and generally argue why these interacting TCIs are beyond either free-fermion TCIs or bosonic SPT phases. 

First we review the logic to classify and construct bosonic SPT phases with both onsite ($G_0$) and crystalline ($G_c$) symmetries\cite{Thorngren2018,Jiang2017,Rasmussen2018,Shiozaki2018,Else2018,Song2018a}. All SPT phases protected by both onsite and crystalline symmetries can be constructed by stacking lower-dimensional SPT phases in a pattern that preserves the crystalline symmetry\cite{Song2017a,Huang2017b,Rasmussen2018,Else2018,Song2018a}. In this work we will focus on a simpler case where the total symmetry group $G=G_0\times G_c$ is a direct product of $G_0$ and $G_c$. For this case, the group cohomology classification of bosonic SPT phases can be decomposed using the K\"unneth formula\cite{Chen2013}
\bea
\notag&\gpc{d+1}{G_c^\ast\times G_0}{U(1)}=\gpc{d+1}{G_c^\ast}{U(1)}\\
\notag&\oplus\gpc{d+1}{G_c^\ast}{\gpc{1}{G_0}{\mbz}}\\
&\bigoplus_{k=0}^{d}\gpc{k}{G_c^\ast}{{\gpc{d-k+1}{G_0}{U(1)}}}.\label{kunneth formula}
\eea
Each term $\gpc{k}{G_c^\ast}{{\gpc{d-k+1}{G_0}{U(1)}}}$ provides a roadmap to construct $d$-dimensional $G$-SPT phases using $(d-k)$-dimensional $G_0$-SPT phases. In particular, not all $(d-k)$-dim. $G_0$-SPT phases are compatible with crystalline symmetry $G_c$\cite{Rasmussen2018}: only the compatible ones are elements of cohomology $\gpc{k}{G_c^\ast}{{\gpc{d-k+1}{G_0}{U(1)}}}$. Moreover, each term of the K\"unneth expansion can be physically realized using the decorated domain wall picture \cite{Chen2014}, allowing an explicit construction. 

\begin{figure}
    \centering
    \includegraphics[width=150pt]{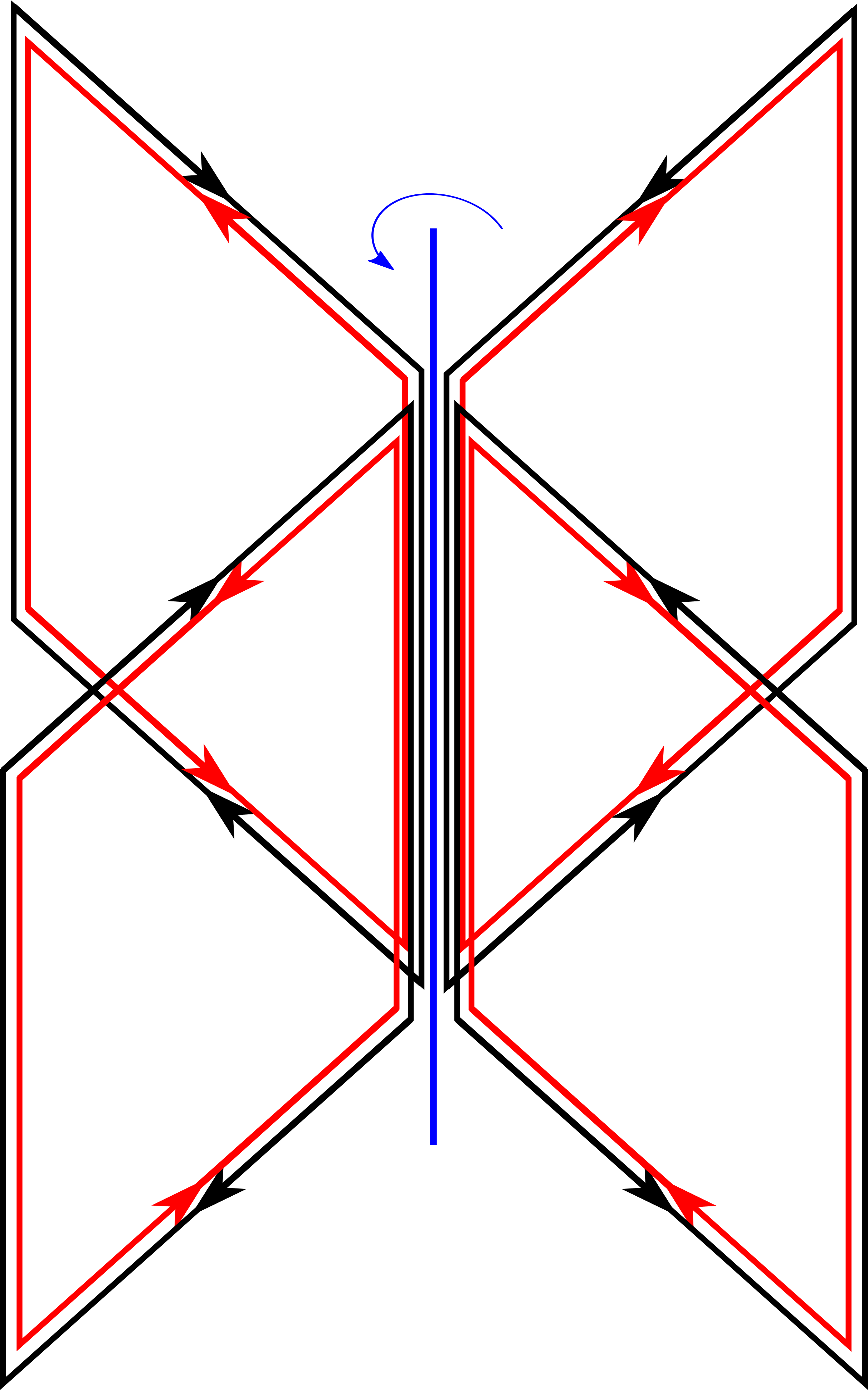}
    \caption{2nd-order 3d interacting TCI with onsite $G_0=U(1)_\text{charge}\times Z_2^{\sigma_z}$ and $C_4$ rotation symmetry. It is constructed by decorating each $C_4$ domain wall by a 2d fermionic TI with a pair of helical edge states protected by $G_0=U(1)_\text{charge}\times Z_2^{\sigma_z}$ symmetry. It is characterized by helical hinge states on each of the 4 hinges of an open system.}
    \label{fig:hinge_tci}
\end{figure}

To be concrete, we consider rotation symmetry $G_c=C_n$ for an example, where SPT phases classified by $\gpc{1}{C_n}{\gpc{d}{G_0}{U(1)}}$ are constructed by stacking $(d-1)$-dimensional $G_0$-SPT phases on the $C_n$ ``domain walls'' as shown in Fig. \ref{fig:plane_tci},\ref{fig:hinge_tci}. Since these $G_0$-SPTs intersect at the $C_n$ rotation axis, the $n$ copies of $G_0$-SPT boundary states must be symmetrically gapped out to ensure a trivial gapped bulk. This requires $n$ copies of the $G_0$-SPT phase to add up to a trivial phase, a condition captured exactly by group cohomology $\gpc{1}{C_n}{\gpc{d}{G_0}{U(1)}}$. 

Although fermion SPT phases are generally beyond the description of group cohomology\cite{Gu2014,Kapustin2015,Lan2018}, the above dimensional reduction construction based on decorated domain wall picture remains valid. Take $C_n$ symmetry for example, instead of using lower-dimensional bosonic SPT phases, we decorate each $C_n$ domain wall by a $(d-1)$-dimensional fermion $G_0$-SPT phase. In particular if the free-fermion classification for symmetry $G_0$ has an integer classification, no free-fermion TCIs can be obtained by decorating $C_n$ domain walls since $\gpc{1}{C_n}{\mathcal{C}_\text{free}^{G_0}\simeq\mbz}=0$. However if interaction reduces the free-fermion integer classification to a finite $\mathcal{C}_\text{int}^{G_0}=\mbz_a$, it is possible to gap out the edge states at the rotation axis by interaction since $\gpc{1}{C_n}{\mathcal{C}_\text{int}^{G_0}\simeq\mbz_a}=\mbz_{(n,a)}$, where $(n,a)$ is the greatest common divisor of integers $n$ and $a$. If the fermion $G_0$-SPT phase on each $C_n$ domain wall cannot be adiabatically tuned into a bosonic $G_0$-SPT phase, we have realized an intrinsically interacting TCI, which is beyond free-fermion TCIs and bosonic SPT phases. 

In the following, we will use this logic to construct interacting fermionic TCIs in two (2d) and three (3d) spatial dimensions. They include 3 examples: 2d and 3d TCIs with $C_4$ rotation symmetry, and 3d TCI with $T_h$ point group symmetry.

\section{2nd-order interacting TCI in $d=2$}\label{sec:k=1,d=2}

In the first example, we consider a two-dimensional (2d) TCI of symmetry class AIII, preserving onsite symmetry
\bea\label{onsite:U(1)xZ2T}
G_0=U(1)\times Z_2^\bst
\eea
and point group $G_c=C_4$ symmetry. This can be realized either in a superconductor with $U(1)_{S_z}$ spin conservation and time reversal $\bst$, a convention we adopt here, or in a TI with an anti-unitary particle-hole symmetry $\bst$.

\begin{figure}
    \centering
    \includegraphics[width=150pt]{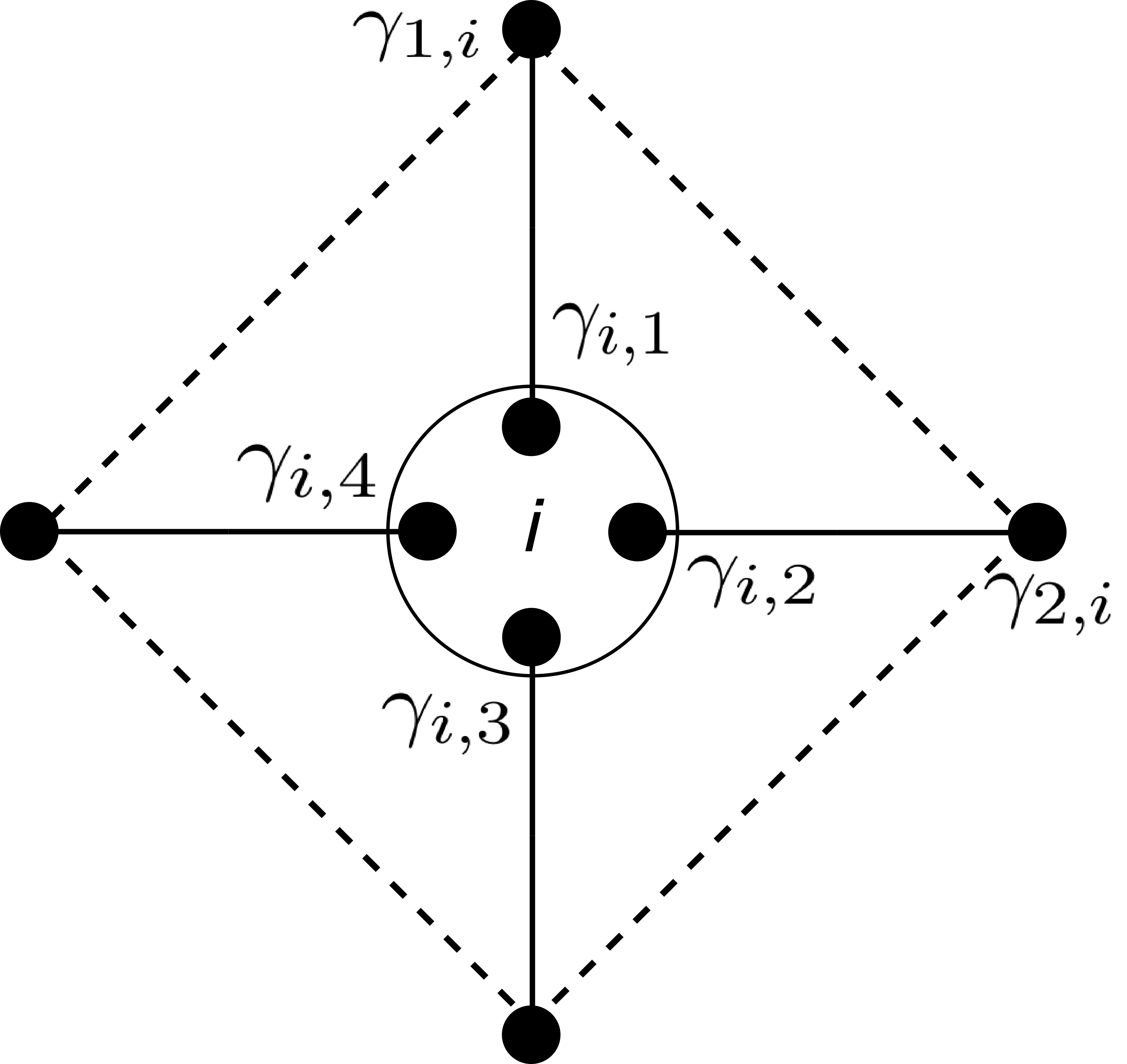}
    \caption{2nd-order 2d interacting TCI, with onsite symmetry (\ref{onsite:U(1)xZ2T}) and $C_4$ rotation symmetry, constructed by stacking 1d $\nu=1$ fermion TI in class AIII on each $C_4$ domain wall. It is characterized by a gapless (complex) fermion mode at each of the 4 corners of an open system.} 
    \label{fig:plane_tci}
\end{figure}

In our minimal model, there is a 4-dimensional Hilbert space of spin-$1/2$ fermions $\{c_{\expval{ij},\alpha}|\alpha=\uparrow,\downarrow\}$ on each nearest-neighbor (NN) link $\expval{ij}$ of the square lattice. In addition to $U(1)_{S_z}$ spin conservation, the time reversal symmetry $\bst$ is also preserved
\bea
c_{\expval{ij},\alpha}\overset{\bst}\longrightarrow\alpha c_{\expval{ij},-\alpha},~~~\alpha=\pm1~\text{for spin}~\uparrow/\downarrow
\eea
Next we reorganize the Hilbert space by writing down a different set of two fermion modes:
\bea
\gamma_{i,j}=\frac1{\sqrt{2}}\big[c_{\expval{ij},\uparrow}+(-1)^i\imth c^\dagger_{\expval{ij},\downarrow}\big],\\
\gamma_{j,i}=\frac1{\sqrt{2}}\big[c_{\expval{ij},\uparrow}+(-1)^j\imth c^\dagger_{\expval{ij},\downarrow}\big]\neq\gamma_{i,j}\label{AIII:end state}
\eea
where we defined sign factor $(-1)^i$ for each lattice site
\bea
(-1)^i\equiv(-1)^{x+y},~~~\vec i=(x,y),~~~x,y\in\mbz.
\eea
As shown in Fig. \ref{fig:plane_tci}, we assign fermion $\gamma_{i,j}$ to site $i$ and $\gamma_{j,i}$ to site $j$. It's straightforward to show they transform under $\bst$ differently depending on the sublattice:
\bea
\gamma_{i,j}\overset{\bst}\longrightarrow(-1)^i\imth\gamma_{i,j}^\dagger,\\
e^{\imth\theta\hat S_z}\gamma_{i,j}e^{-\imth\theta\hat S_z}=e^{-\imth\frac{\theta}{2}}\gamma_{i,j}
\eea
The following interacting Hamiltonian
\bea\label{ham:2d:AIII}
&\hat H^{2d}_{C_4,\text{AIII}}=\sum_i\hat H(i),\\
\notag&\hat H(i)=U\sum_{j=1}^4(\gamma^\dagger_{i,j}\gamma_{i,j}-\frac12)(\gamma^\dagger_{i,j+1}\gamma_{i,j+1}-\frac12)\\
\notag&+J(\imth\gamma_{i,1}^\dagger\gamma_{i,2}\gamma^\dagger_{i,3}\gamma_{i,4}+~h.c.),~~~U,J>0\\
&\notag 1\leq j\leq4\in\text{NN}(i)~\text{as shown in Fig. \ref{fig:plane_tci}.}
\eea
can gap out all fermions symmetrically, resulting in a TCI ground state. As illustrated in Fig. \ref{fig:plane_tci}, this TCI features a gapless fermion $\gamma_{j,i}$ at each corner $1\leq j\leq4$ of an open system, protected by onsite symmetry (\ref{onsite:U(1)xZ2T}). 

First we show this is an interacting TCI that cannot be realized by free fermion Hamiltonians. 1d free-fermion TIs in class AIII have a $\nu\in\mbz$ classification, where the 0d end state of $\nu=1$ TI is nothing but a complex fermion mode $\gamma_{j,i}$ in (\ref{AIII:end state}). When we compactify the open system to achieve a closed system with periodic boundary condition (PBC), the 4 gapless corner modes must be gapped out symmetrically to recover a gapped bulk. However free-fermion integer classification dictates that it's impossible to gap out these 4 copies of $\nu=1$ end state by any free-fermion Hamiltonian. Therefore our model (\ref{ham:2d:AIII}) is beyond any free-fermion TCIs. As shown by \Ref{Tang2012a}, the 1d free-fermion $\mbz$ classification in class AIII is reduced to a $\mbz_4$ classification for symmetry (\ref{onsite:U(1)xZ2T}), enabling us to construct model (\ref{ham:2d:AIII}) to symmetrically gap out where 4 copies of $\nu=1$ end state. 

Next we show this interacting TCI is also beyond the framework of bosonic SPT phases. As shown in \Ref{Rasmussen2018}, 2nd-order bosonic SPT phases with $G=C_4\times G_0$ in 2d are classified by 
\bea
\mbz_2^2=\gpc{1}{C_4\simeq Z_4}{\gpc{2}{G_0}{U(1)}}=\gpc{1}{C_4\simeq Z_4}{\mbz_2^2}
\eea
These 2nd-order SPT phases are constructed by decorating each $C_4$ domain wall with a 1d bosonic $G_0$-SPT phase, and their corner states are the end state of 1d boson SPT phases. However the corner fermion mode $\gamma_{j,i}$ in (\ref{AIII:end state}) can never to realized at the end of a 1d boson SPT phase, and therefore model (\ref{ham:2d:AIII}) hosts an intrinsically interacting TCI of fermions. 

Lastly, we comment that this intreacting TCI remains stable even if the onsite symmetry (\ref{onsite:U(1)xZ2T}) is broken down to a subgroup $G_0=Z_2^\bst$ with $\bst^2=+1$, where each $\nu=1$ 1d TI in class AIII is now reduced to a $\nu=2$ 1d topological superconductor in class BDI.

\section{2nd-order interacting TCI in $d=3$}\label{sec:k=1,d=3}

Next we consider 3d TCI preserving $G_c=C_4$ rotational symmetry, and an onsite symmetry
\bea
\label{onsite:U(1)xZ2}
G_0=U(1)_\text{charge}\times Z_2^{\sigma_z}
\eea

The building block of our intrinsically interacting TCI is a $\nu=1$ 2d fermion TI protected by onsite symmetry (\ref{onsite:U(1)xZ2}), with a pair of helical edge modes of opposite $S_z$ quantum number $s=\pm1$:
\bea\label{helical edge}
\mathcal{L}_0=\sum_{s-\pm1}\psi_s^\dagger(\imth\partial_t-s\cdot v\partial_x)\psi_s\\
\psi_s\overset{e^{\imth\theta\hat Q}}\longrightarrow e^{-\imth\theta}\psi,~~~\psi_s\overset{\sigma_z}\longrightarrow s\cdot\psi_s.
\eea
It can be easily realized in \eg Kane-Mele model\cite{Kane2005}. The above helical edge modes can be bosonized as
\bea
\psi_s\sim\eta_s e^{\imth\phi_s},~~[\phi_{s_1}(x),\phi_{s_2}(y)]=\imth s_1\pi\text{Sgn}(x-y)\delta_{s_1,s_2}~~
\eea
where $\eta_s$ are Klein factors and $\{\phi_s\}$ are chiral bosons.

As shown in Fig. \ref{fig:hinge_tci}, each of the four $C_4$ ``domain walls'' is decorated by such a 2d TI, where the four helical edge modes $\{\phi_s^a|1\leq a\leq4\}$ intersect at the $C_4$ rotation axis. The symmetries are implemented as follows:
\bea
\phi_s^a\overset{e^{\imth\theta\hat Q}}\longrightarrow\phi_s^a-\theta,\\
\phi_s^a\overset{\sigma_z}\longrightarrow\phi_s^a+\frac{1-s}{2}\pi,\\
\phi_s^a\overset{C_4}\longrightarrow\phi_s^{a+1}. 
\eea
Next we construct a fully symmetric interacting Hamiltonian that gap out these 4 helical edge states on the $C_4$ axis. First we consider 
\bea
\hat H_1=-V_1\sum_{a=1}^4\cos(\phi_+^a+\phi_+^{a+2}-\phi_-^{a+2}-\phi_-^{a})
\eea
which already gaps out four chiral bosons, leaving the gapless modes below
\bea
\varphi_a &\equiv\phi^a_+-\phi^a_-\sim\phi_-^{a+2}-\phi_+^{a+2},~~~a=1,2;\\
\theta_a &\equiv\phi^a_+-\phi^{a+2}_-\sim\phi_-^{a}-\phi_+^{a+2},~~~a=1,2.
\eea
which transform under symmetries as
\bea
\bpm\varphi_1\\ \theta_1\epm\overset{C_4}\longrightarrow
\bpm\varphi_2\\ \theta_2\epm\overset{C_4}\longrightarrow-\bpm\varphi_1\\ \theta_1\epm.
\eea
Therefore we can write down the following symmetric Hamiltonian
\bea\label{ham:3d:C4}
&\hat H_\text{int}=\hat H_0+\hat H_1+\hat H_2,\\
\notag&\hat H_2=-V_2\big[\cos(\varphi_1+\theta_2)+\cos(\varphi_2-\theta_1)\big]
\eea
where $\hat H_0$ is a free-fermion lattice model realizing the Kane-Mele model on each $C_4$ domain wall. $\hat H_\text{int}$ symmetrically gaps out all the helical edge modes on the $C_4$ axis, leading to a TCI ground state. Meanwhile notice that this is a 2nd-order fermion SPT phase characterized by 4 gapless $\nu=1$ helical hinge modes in an open system, which are robust against any weak symmetric perturbations.

To see that this TCI cannot be realized by a free-fermion Hamiltonian, we compactify the open system on a three torus, where the 4 gapless hinge modes are joined together after compactification. Within free fermion Hamiltonians, there is no way to symmetrically gap out these 4 helical modes due to the $\nu\in\mbz$ classification of free fermions with symmetry (\ref{onsite:U(1)xZ2}). However as shown in \Ref{Qi2013,Ryu2012,Yao2013,Lu2012a}, interactions reduce the free-fermion $\mbz$ classification to a finite $\mbz_4$ classification, allows 4 copies of helical edge states to gap out while preserving symmetry (\ref{onsite:U(1)xZ2}). This reveals why the $\nu=1$ 2d TI with onsite symmetry (\ref{onsite:U(1)xZ2}) is compatible with $C_4$ rotational symmetry to construct this interacting 2nd-order TCI in 3d. 
 
Finally we show that this interacting TCI is not a bosonic SPT phase. As shown in \Ref{Rasmussen2018}, 2nd-order 3d bosonic SPT phases with symmetry group $G=C_4\times G_0$ are classified by 
\bea
\mbz_2^2=\gpc{1}{C_4}{\gpc{3}{G_0}{U(1)}}=\gpc{1}{Z_4}{\mbz\times\mbz_2^2}
\eea
They are characterized by gapless 1d hinge states which are edge modes of 2d bosonic $G_0$-SPT phases. Since helical edge modes (\ref{helical edge}) cannot be realized in any 2d bosonic SPT phases\cite{Lu2012a}, the ground state of model (\ref{ham:3d:C4}) cannot be a boson SPT phase. Therefore we have shown that model (\ref{ham:3d:C4}) realizes an intrinsically interacting TCI of fermions. 

\section{3rd-order interacting TCI in $d=3$}\label{sec:k=2,d=3}

Lastly we consider 3d superconductors in symmetry class BDI, with time reversal symmetry $G_0=Z_2^\bst$ satisfying $\bst^2=+1$. Below we construct an intrinsically interacting TCI with both time reversal and pyritohedral  point group symmetry $G_c=T_h$.

\begin{figure}
    \centering
    \includegraphics[width=200pt]{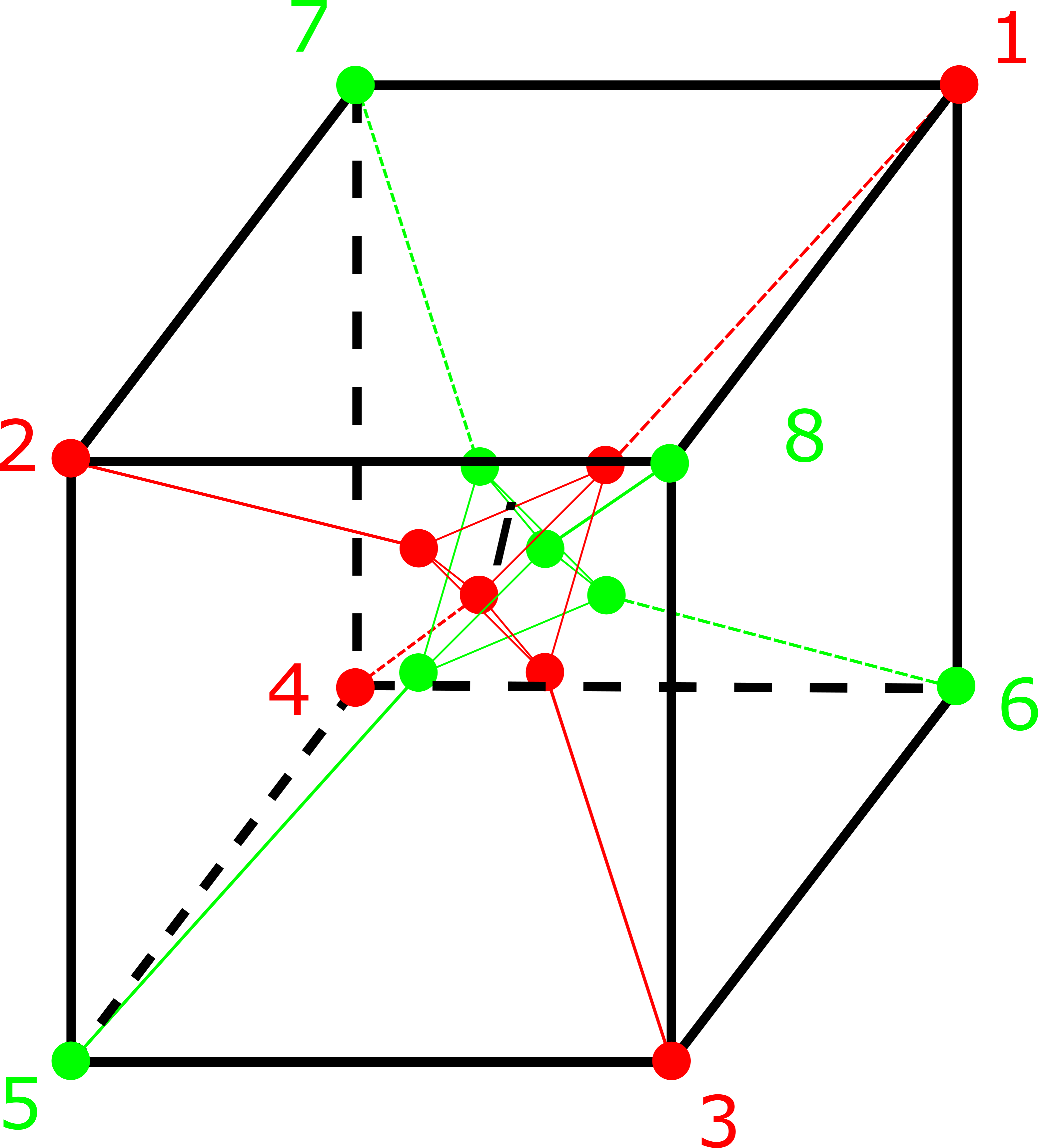}
    \caption{3rd-order interacting TCI with onsite $Z_2^\bst$ ($\bst^2=+1$) and pyritohedral  point group symmetry $G_c=T_h$. It is constructed by stacking 1d $\nu=1$ Kitaev chain in class BDI on the eight 3-fold rotation axes labeled by green and red lines, and characterized by a single Majorana zero modes at each of the eight corners on a cube-shaped open system.}
    \label{fig:3d_tci}
\end{figure}

We consider a lattice model of spinless fermions, where one single fermion mode $c_{\expval{i,j}}$ lives at the center of each nearest-neighbor (NN) link $\expval{i,j}$ of the body-centered cubic (BCC) lattice:
\bea
\vec i=i_1(-1,1,1)+i_2(1,-1,1)+i_3(1,1,-1),~~~i_{1,2,3}\in\mbz.\notag
\eea
As shown in Fig. \ref{fig:3d_tci}, the NN link centers form a cubic lattice. The complex fermion on each NN link can be represented by two Majorana fermions $\{\chi_{i,j}\}$ living on the sites $\{i\}$ of the BCC lattice (or centers of the cube in Fig. \ref{fig:3d_tci}):
\bea
&c_{\expval{i,j}}=(\chi_{i,j}+\imth\chi_{j,i})/{2},\\
&\notag i_1+i_2+i_3=0\mod2,~~j_1+j_2+j_3=1\mod2. 
\eea
where we have chosen $i$ and $\chi_{i,j}$ to live on the even sublattice, $j$ and $\chi_{j,i}$ on the odd sublattice of the BCC lattice. Under time reversal symmetry the fermions transform as
\bea\label{3d:BDI:sym:trs}
c_{\expval{i,j}}\overset{\bst}\longrightarrow c_{\expval{i,j}},~~~\chi_{i,j}\overset{\bst}\longrightarrow(-1)^i\chi_{i,j}
\eea
where we defined sign $(-1)^i\equiv(-1)^{i_1+i_2+i_3}$ for each site $i$. The point group $T_h$ is generated by 3-fold rotation $R_3$, 2-fold rotation $R_{x}$ and inversion $\bsi$:
\bea
(x,y,z)\overset{R_3}\longrightarrow(y,z,x),\\
(x,y,z)\overset{R_x}\longrightarrow(x,-y,-z),\\
(x,y,z)\overset{\bsi}\longrightarrow(-x,-y,-z).
\eea
In Fig. \ref{fig:3d_tci} we label the eight NNs of an even site $i$ at the cube center (also the inversion center) as $1\leq j\leq 8$, and the 3-fold $R_3$ axis crosses sites 1, 5 and cube center $i$. The eight Majoranas $\{\chi_{i,j}|1\leq j\leq 8\}$ living on site $i$ transform under inversion as
\bea
\chi_{i,a}\overset{\bsi}\longleftrightarrow\chi_{i,a+4},~~~1\leq a\leq4.
\eea
In Fig. \ref{fig:3d_tci} we color $1\leq j\leq 4$ in red and $5\leq j\leq8$ in green. Under 3-fold and 2-fold rotations the 4 red Majoranas transform as
\bea
\chi_{i,1}\overset{R_3}\longrightarrow\chi_{i,1},~~\chi_{i,2}\overset{R_3}\longrightarrow\chi_{i,3}\overset{R_3}\longrightarrow\chi_{i,4}\overset{R_3}\longrightarrow\chi_{i,2},\\
\chi_{i,1}\overset{R_x}\longleftrightarrow\chi_{i,3},~~~\chi_{i,2}\overset{R_x}\longleftrightarrow\chi_{i,4}.
\eea
and similarly for the 4 green ones. 

The fermion TCI is obtained by the following symmetric interacting Hamiltonian
\bea\label{ham:3d:BDI}
&\hat H^{3d}_{T_h,\text{BDI}}=\sum_i\hat H(i),\\
\notag&\hat H(i)=U(\chi_{i,1}\chi_{i,2}\chi_{i,3}\chi_{i,4}+\chi_{i,5}\chi_{i,6}\chi_{i,7}\chi_{i,8})\\
\notag&+J~\hat P_{U}\chi_{i,1}\chi_{i,5}(\chi_{i,2}\chi_{i,6}+\chi_{i,3}\chi_{i,7}+\chi_{i,4}\chi_{i,8})\hat P_U,\\
&\hat P_U\equiv\prod_i\frac{1-\chi_{i,1}\chi_{i,2}\chi_{i,3}\chi_{i,4}}2\cdot\frac{1-\chi_{i,5}\chi_{i,6}\chi_{i,7}\chi_{i,8}}2,\notag\\
&U,J>0,~~~\notag 1\leq j\leq 8\in\text{NN}(i)~\text{see Fig. \ref{fig:3d_tci}}.
\eea
When projected (by projector $\hat P_U$) into the ground state manifold $\chi_{i,1}\chi_{i,2}\chi_{i,3}\chi_{i,4}=\chi_{i,1}\chi_{i,2}\chi_{i,3}\chi_{i,4}=-1$ of the $U$ term, the $J$ term is nothing but an antiferromagnetic Heisenberg interaction between the effective spin-$1/2$ of 4 red Majoranas and the spin-$1/2$ of 4 green Majoranas. Therefore Hamiltonian (\ref{ham:3d:BDI}) has a unique gapped ground state which preserves all symmetries. 

Below we show this ground state of (\ref{ham:3d:BDI}) is an intrinsically interacting TCI. First as illustrated in Fig. \ref{fig:3d_tci}, there will be a single Majorana zero mode (MZM) at each corner of a cubic-shaped open system, which is robust against any perturbations. 3d bosonic $G=Z_2^\bst\times T_h$-SPT phases with robust corner states classified in \Ref{Rasmussen2018} using group cohomology formula
\bea
\gpc{2}{T_h^\ast}{\gpc{2}{Z_2^\bst}{U(1)}}=\mbz_2^3
\eea
They are constructed by decorating 3-fold and 2-fold rotational axes by 1d $Z_2^\bst$-protected Haldane chain, characterized by a Kramers doublet (\ie spin-$1/2$) at each corner. Therefore our model (\ref{ham:3d:BDI}) hosting Majorana corner modes is clearly beyond bosonic SPT phases. 

Meanwhile as shown in (\ref{3d:BDI:sym:trs}), the 8 corner MZMs $\{\chi_{a|1\leq a\leq8}\}$ of the open system are all even (or odd) under time reversal symmetry $\bst$. Now let us glue the open boundary of the finite cubic-shaped system into a closed system with the periodic boundary conditions. In particular, the 8 MZMs must be gapped out symmetrically to recover a gapped bulk.. However, any bilinear coupling $\imth\chi_a\chi_b$ is forbidden by time reversal symmetry. Therefore it is impossible to recover a gapped bulk within the space of free-fermion Hamiltonians, and we have proved by contradiction that the ground state of model (\ref{ham:3d:BDI}) cannot be adiabatically connected to any free fermion Hamiltonian without closing the gap. 

Therefore model (\ref{ham:3d:BDI}) realizes neither a free-fermion TCI nor a bosonic SPT phase, but an intrinsically interacting TCI of fermions. Pictorially, this interacting TCI is constructed by decorating each of the 8 $R_3$ axes in Fig. \ref{fig:3d_tci} by a $\nu=1$ Kitaev chain\cite{Kitaev2001} in symmetry class BDI (with $\bst^2=+1$). The 8 Kitaev chains terminate and intersect at the inversion center $i$, giving rise to 8 MZMs at each site $i$. As shown by Fidkowski and Kitaev\cite{Fidkowski2010}, interactions reduce the free-fermion integer classification $\nu\in\mbz$ of 1d class BDI to a $\mbz_8$ classification, where 8 MZMs with time reversal symmetry (\ref{3d:BDI:sym:trs}) can be gapped out symmetrically. This is exactly what model (\ref{ham:3d:BDI}) achieved.

\section{Discussions}\label{sec:discussion}

To summarize, we provide a general construction for intrinsically interacting TCIs of fermions, which are beyond the description of either free-fermion Hamiltonians and interacting boson SPT phases. We use three explicit examples in two and three dimensions to demonstrate this construction, and show that these interacting TCIs are often characterized by robust fermion modes on corners/hinges of an open system. 

Generally in a fermion system, the fermion symmetry group $G_f$ is an central extension of the physical symmetry group $G$ by the fermion parity $Z_2^f\equiv\{1,(-1)^{\hat F}\}$, \ie $G_f/Z_2^f=G$. In all 3 examples we presented, $G_f$ has a direct-product form:
\bea
G_f=G\times Z_2^f=G_0\times G_c\times Z_2^f.  
\eea
However this is not always the case for a generic TCI in the decorated domain wall construction. For example in Fig. \ref{fig:plane_tci}, once we replace the $\nu=1$ TCI in class AIII on each $C_4$ domain wall by a 1d Kitaev chain, it is impossible to gap out the 4 MZMs at $C_4$ center if $(C_4)^4=1$, since the $C_4$ operation will change the total fermion parity of the 4 MZMs. Meanwhile, a nontrivial extension by fermion parity $(C_4)^4=(-1)^{\hat F}$ is compatible with a gapped bulk. Though not discussed in our work, such a nontrivial interplay between onsite and crystalline symmetries can be important to understand a full classification of fermonic SPT phases with both onsite and crystalline symmetries.

\acknowledgments

YML thanks Aspen Center for Physics for hospitality, where this work was initiated. This work is supported
by NSF under award number DMR-1653769 (AR,YML),
and in part by NSF grant PHY-1607611 (YML). 

Upon completion of this work, we became aware of a related work by Meng Cheng and Chenjie Wang who classified fermionic SPT phases with rotational symmetry. Their work will appear on arXiv on the same date with our work.

\bibliographystyle{apsrev4-1}
\bibliography{bibs_career}

\begin{thebibliography}{41}%
\makeatletter
\providecommand \@ifxundefined [1]{%
 \@ifx{#1\undefined}
}%
\providecommand \@ifnum [1]{%
 \ifnum #1\expandafter \@firstoftwo
 \else \expandafter \@secondoftwo
 \fi
}%
\providecommand \@ifx [1]{%
 \ifx #1\expandafter \@firstoftwo
 \else \expandafter \@secondoftwo
 \fi
}%
\providecommand \natexlab [1]{#1}%
\providecommand \enquote  [1]{``#1''}%
\providecommand \bibnamefont  [1]{#1}%
\providecommand \bibfnamefont [1]{#1}%
\providecommand \citenamefont [1]{#1}%
\providecommand \href@noop [0]{\@secondoftwo}%
\providecommand \href [0]{\begingroup \@sanitize@url \@href}%
\providecommand \@href[1]{\@@startlink{#1}\@@href}%
\providecommand \@@href[1]{\endgroup#1\@@endlink}%
\providecommand \@sanitize@url [0]{\catcode `\\12\catcode `\$12\catcode
  `\&12\catcode `\#12\catcode `\^12\catcode `\_12\catcode `\%12\relax}%
\providecommand \@@startlink[1]{}%
\providecommand \@@endlink[0]{}%
\providecommand \url  [0]{\begingroup\@sanitize@url \@url }%
\providecommand \@url [1]{\endgroup\@href {#1}{\urlprefix }}%
\providecommand \urlprefix  [0]{URL }%
\providecommand \Eprint [0]{\href }%
\providecommand \doibase [0]{http://dx.doi.org/}%
\providecommand \selectlanguage [0]{\@gobble}%
\providecommand \bibinfo  [0]{\@secondoftwo}%
\providecommand \bibfield  [0]{\@secondoftwo}%
\providecommand \translation [1]{[#1]}%
\providecommand \BibitemOpen [0]{}%
\providecommand \bibitemStop [0]{}%
\providecommand \bibitemNoStop [0]{.\EOS\space}%
\providecommand \EOS [0]{\spacefactor3000\relax}%
\providecommand \BibitemShut  [1]{\csname bibitem#1\endcsname}%
\let\auto@bib@innerbib\@empty
%</preamble>
\bibitem [{\citenamefont {Hasan}\ and\ \citenamefont {Kane}(2010)}]{Hasan2010}%
  \BibitemOpen
  \bibfield  {author} {\bibinfo {author} {\bibfnamefont {M.~Z.}\ \bibnamefont
  {Hasan}}\ and\ \bibinfo {author} {\bibfnamefont {C.~L.}\ \bibnamefont
  {Kane}},\ }\href {http://link.aps.org/doi/10.1103/RevModPhys.82.3045}
  {\bibfield  {journal} {\bibinfo  {journal} {Rev. Mod. Phys.}\ }\textbf
  {\bibinfo {volume} {82}},\ \bibinfo {pages} {3045} (\bibinfo {year}
  {2010})}\BibitemShut {NoStop}%
\bibitem [{\citenamefont {Hasan}\ and\ \citenamefont
  {Moore}(2011)}]{Hasan2011}%
  \BibitemOpen
  \bibfield  {author} {\bibinfo {author} {\bibfnamefont {M.~Z.}\ \bibnamefont
  {Hasan}}\ and\ \bibinfo {author} {\bibfnamefont {J.~E.}\ \bibnamefont
  {Moore}},\ }\bibfield  {booktitle} {\emph {\bibinfo {booktitle} {Annual
  Review of Condensed Matter Physics}},\ }\href
  {http://dx.doi.org/10.1146/annurev-conmatphys-062910-140432} {\bibfield
  {journal} {\bibinfo  {journal} {Annu. Rev. Condens. Matter Phys.}\ }\textbf
  {\bibinfo {volume} {2}},\ \bibinfo {pages} {55} (\bibinfo {year}
  {2011})}\BibitemShut {NoStop}%
\bibitem [{\citenamefont {Qi}\ and\ \citenamefont {Zhang}(2011)}]{Qi2011c}%
  \BibitemOpen
  \bibfield  {author} {\bibinfo {author} {\bibfnamefont {X.-L.}\ \bibnamefont
  {Qi}}\ and\ \bibinfo {author} {\bibfnamefont {S.-C.}\ \bibnamefont {Zhang}},\
  }\href {http://link.aps.org/doi/10.1103/RevModPhys.83.1057} {\bibfield
  {journal} {\bibinfo  {journal} {Rev. Mod. Phys.}\ }\textbf {\bibinfo {volume}
  {83}},\ \bibinfo {pages} {1057} (\bibinfo {year} {2011})}\BibitemShut
  {NoStop}%
\bibitem [{\citenamefont {Chen}\ \emph {et~al.}(2013)\citenamefont {Chen},
  \citenamefont {Gu}, \citenamefont {Liu},\ and\ \citenamefont
  {Wen}}]{Chen2013}%
  \BibitemOpen
  \bibfield  {author} {\bibinfo {author} {\bibfnamefont {X.}~\bibnamefont
  {Chen}}, \bibinfo {author} {\bibfnamefont {Z.-C.}\ \bibnamefont {Gu}},
  \bibinfo {author} {\bibfnamefont {Z.-X.}\ \bibnamefont {Liu}}, \ and\
  \bibinfo {author} {\bibfnamefont {X.-G.}\ \bibnamefont {Wen}},\ }\href
  {http://link.aps.org/doi/10.1103/PhysRevB.87.155114} {\bibfield  {journal}
  {\bibinfo  {journal} {Phys. Rev. B}\ }\textbf {\bibinfo {volume} {87}},\
  \bibinfo {pages} {155114} (\bibinfo {year} {2013})}\BibitemShut {NoStop}%
\bibitem [{\citenamefont {Senthil}(2015)}]{Senthil2015}%
  \BibitemOpen
  \bibfield  {author} {\bibinfo {author} {\bibfnamefont {T.}~\bibnamefont
  {Senthil}},\ }\bibfield  {booktitle} {\emph {\bibinfo {booktitle} {Annual
  Review of Condensed Matter Physics}},\ }\href {\doibase
  10.1146/annurev-conmatphys-031214-014740} {\bibfield  {journal} {\bibinfo
  {journal} {Annu. Rev. Condens. Matter Phys.}\ }\textbf {\bibinfo {volume}
  {6}},\ \bibinfo {pages} {299} (\bibinfo {year} {2015})}\BibitemShut {NoStop}%
\bibitem [{\citenamefont {Schnyder}\ \emph {et~al.}(2008)\citenamefont
  {Schnyder}, \citenamefont {Ryu}, \citenamefont {Furusaki},\ and\
  \citenamefont {Ludwig}}]{Schnyder2008}%
  \BibitemOpen
  \bibfield  {author} {\bibinfo {author} {\bibfnamefont {A.~P.}\ \bibnamefont
  {Schnyder}}, \bibinfo {author} {\bibfnamefont {S.}~\bibnamefont {Ryu}},
  \bibinfo {author} {\bibfnamefont {A.}~\bibnamefont {Furusaki}}, \ and\
  \bibinfo {author} {\bibfnamefont {A.~W.~W.}\ \bibnamefont {Ludwig}},\ }\href
  {http://link.aps.org/doi/10.1103/PhysRevB.78.195125} {\bibfield  {journal}
  {\bibinfo  {journal} {Phys. Rev. B}\ }\textbf {\bibinfo {volume} {78}},\
  \bibinfo {pages} {195125} (\bibinfo {year} {2008})}\BibitemShut {NoStop}%
\bibitem [{\citenamefont {Kitaev}(2009)}]{Kitaev2009}%
  \BibitemOpen
  \bibfield  {author} {\bibinfo {author} {\bibfnamefont {A.}~\bibnamefont
  {Kitaev}},\ }\href {http://link.aip.org/link/?APC/1134/22/1} {\bibfield
  {journal} {\bibinfo  {journal} {AIP Conf. Proc.}\ }\textbf {\bibinfo {volume}
  {1134}},\ \bibinfo {pages} {22} (\bibinfo {year} {2009})}\BibitemShut
  {NoStop}%
\bibitem [{\citenamefont {Chiu}\ \emph {et~al.}(2016)\citenamefont {Chiu},
  \citenamefont {Teo}, \citenamefont {Schnyder},\ and\ \citenamefont
  {Ryu}}]{Chiu2016}%
  \BibitemOpen
  \bibfield  {author} {\bibinfo {author} {\bibfnamefont {C.-K.}\ \bibnamefont
  {Chiu}}, \bibinfo {author} {\bibfnamefont {J.~C.~Y.}\ \bibnamefont {Teo}},
  \bibinfo {author} {\bibfnamefont {A.~P.}\ \bibnamefont {Schnyder}}, \ and\
  \bibinfo {author} {\bibfnamefont {S.}~\bibnamefont {Ryu}},\ }\href
  {https://link.aps.org/doi/10.1103/RevModPhys.88.035005} {\bibfield  {journal}
  {\bibinfo  {journal} {Rev. Mod. Phys.}\ }\textbf {\bibinfo {volume} {88}},\
  \bibinfo {pages} {035005} (\bibinfo {year} {2016})}\BibitemShut {NoStop}%
\bibitem [{\citenamefont {{Kapustin}}(2014)}]{Kapustin2014}%
  \BibitemOpen
  \bibfield  {author} {\bibinfo {author} {\bibfnamefont {A.}~\bibnamefont
  {{Kapustin}}},\ }\href@noop {} {\bibfield  {journal} {\bibinfo  {journal}
  {ArXiv e-prints}\ } (\bibinfo {year} {2014})},\ \Eprint
  {http://arxiv.org/abs/1403.1467} {arXiv:1403.1467 [cond-mat.str-el]}
  \BibitemShut {NoStop}%
\bibitem [{\citenamefont {Gu}\ and\ \citenamefont {Wen}(2014)}]{Gu2014}%
  \BibitemOpen
  \bibfield  {author} {\bibinfo {author} {\bibfnamefont {Z.-C.}\ \bibnamefont
  {Gu}}\ and\ \bibinfo {author} {\bibfnamefont {X.-G.}\ \bibnamefont {Wen}},\
  }\href {https://link.aps.org/doi/10.1103/PhysRevB.90.115141} {\bibfield
  {journal} {\bibinfo  {journal} {PRB}\ }\textbf {\bibinfo {volume} {90}},\
  \bibinfo {pages} {115141} (\bibinfo {year} {2014})}\BibitemShut {NoStop}%
\bibitem [{\citenamefont {Kapustin}\ \emph {et~al.}(2015)\citenamefont
  {Kapustin}, \citenamefont {Thorngren}, \citenamefont {Turzillo},\ and\
  \citenamefont {Wang}}]{Kapustin2015}%
  \BibitemOpen
  \bibfield  {author} {\bibinfo {author} {\bibfnamefont {A.}~\bibnamefont
  {Kapustin}}, \bibinfo {author} {\bibfnamefont {R.}~\bibnamefont {Thorngren}},
  \bibinfo {author} {\bibfnamefont {A.}~\bibnamefont {Turzillo}}, \ and\
  \bibinfo {author} {\bibfnamefont {Z.}~\bibnamefont {Wang}},\ }\href
  {https://doi.org/10.1007/JHEP12(2015)052} {\bibfield  {journal} {\bibinfo
  {journal} {Journal of High Energy Physics}\ }\textbf {\bibinfo {volume}
  {2015}},\ \bibinfo {pages} {1} (\bibinfo {year} {2015})}\BibitemShut
  {NoStop}%
\bibitem [{\citenamefont {Kapustin}\ and\ \citenamefont
  {Thorngren}(2017)}]{Kapustin2017}%
  \BibitemOpen
  \bibfield  {author} {\bibinfo {author} {\bibfnamefont {A.}~\bibnamefont
  {Kapustin}}\ and\ \bibinfo {author} {\bibfnamefont {R.}~\bibnamefont
  {Thorngren}},\ }\href {https://doi.org/10.1007/JHEP10(2017)080} {\bibfield
  {journal} {\bibinfo  {journal} {Journal of High Energy Physics}\ }\textbf
  {\bibinfo {volume} {2017}},\ \bibinfo {pages} {80} (\bibinfo {year}
  {2017})}\BibitemShut {NoStop}%
\bibitem [{\citenamefont {Wang}\ and\ \citenamefont {Gu}(2018)}]{Wang2018d}%
  \BibitemOpen
  \bibfield  {author} {\bibinfo {author} {\bibfnamefont {Q.-R.}\ \bibnamefont
  {Wang}}\ and\ \bibinfo {author} {\bibfnamefont {Z.-C.}\ \bibnamefont {Gu}},\
  }\href {https://link.aps.org/doi/10.1103/PhysRevX.8.011055} {\bibfield
  {journal} {\bibinfo  {journal} {PRX}\ }\textbf {\bibinfo {volume} {8}},\
  \bibinfo {pages} {011055} (\bibinfo {year} {2018})}\BibitemShut {NoStop}%
\bibitem [{\citenamefont {Cheng}\ \emph
  {et~al.}(2018{\natexlab{a}})\citenamefont {Cheng}, \citenamefont {Bi},
  \citenamefont {You},\ and\ \citenamefont {Gu}}]{Cheng2018a}%
  \BibitemOpen
  \bibfield  {author} {\bibinfo {author} {\bibfnamefont {M.}~\bibnamefont
  {Cheng}}, \bibinfo {author} {\bibfnamefont {Z.}~\bibnamefont {Bi}}, \bibinfo
  {author} {\bibfnamefont {Y.-Z.}\ \bibnamefont {You}}, \ and\ \bibinfo
  {author} {\bibfnamefont {Z.-C.}\ \bibnamefont {Gu}},\ }\href
  {https://link.aps.org/doi/10.1103/PhysRevB.97.205109} {\bibfield  {journal}
  {\bibinfo  {journal} {PRB}\ }\textbf {\bibinfo {volume} {97}},\ \bibinfo
  {pages} {205109} (\bibinfo {year} {2018}{\natexlab{a}})}\BibitemShut
  {NoStop}%
\bibitem [{\citenamefont {{Lan}}\ \emph {et~al.}(2018)\citenamefont {{Lan}},
  \citenamefont {{Zhu}},\ and\ \citenamefont {{Wen}}}]{Lan2018}%
  \BibitemOpen
  \bibfield  {author} {\bibinfo {author} {\bibfnamefont {T.}~\bibnamefont
  {{Lan}}}, \bibinfo {author} {\bibfnamefont {C.}~\bibnamefont {{Zhu}}}, \ and\
  \bibinfo {author} {\bibfnamefont {X.-G.}\ \bibnamefont {{Wen}}},\ }\href@noop
  {} {\bibfield  {journal} {\bibinfo  {journal} {ArXiv e-prints}\ } (\bibinfo
  {year} {2018})},\ \Eprint {http://arxiv.org/abs/1809.01112} {arXiv:1809.01112
  [cond-mat.str-el]} \BibitemShut {NoStop}%
\bibitem [{\citenamefont {Lu}\ and\ \citenamefont
  {Vishwanath}(2012)}]{Lu2012a}%
  \BibitemOpen
  \bibfield  {author} {\bibinfo {author} {\bibfnamefont {Y.-M.}\ \bibnamefont
  {Lu}}\ and\ \bibinfo {author} {\bibfnamefont {A.}~\bibnamefont
  {Vishwanath}},\ }\href {http://link.aps.org/doi/10.1103/PhysRevB.86.125119}
  {\bibfield  {journal} {\bibinfo  {journal} {Phys. Rev. B}\ }\textbf {\bibinfo
  {volume} {86}},\ \bibinfo {pages} {125119} (\bibinfo {year}
  {2012})}\BibitemShut {NoStop}%
\bibitem [{\citenamefont {Wang}\ and\ \citenamefont {Levin}(2015)}]{Wang2015}%
  \BibitemOpen
  \bibfield  {author} {\bibinfo {author} {\bibfnamefont {C.}~\bibnamefont
  {Wang}}\ and\ \bibinfo {author} {\bibfnamefont {M.}~\bibnamefont {Levin}},\
  }\href {http://link.aps.org/doi/10.1103/PhysRevB.91.165119} {\bibfield
  {journal} {\bibinfo  {journal} {Phys. Rev. B}\ }\textbf {\bibinfo {volume}
  {91}},\ \bibinfo {pages} {165119} (\bibinfo {year} {2015})}\BibitemShut
  {NoStop}%
\bibitem [{\citenamefont {Wang}\ \emph {et~al.}(2017)\citenamefont {Wang},
  \citenamefont {Lin},\ and\ \citenamefont {Gu}}]{Wang2017}%
  \BibitemOpen
  \bibfield  {author} {\bibinfo {author} {\bibfnamefont {C.}~\bibnamefont
  {Wang}}, \bibinfo {author} {\bibfnamefont {C.-H.}\ \bibnamefont {Lin}}, \
  and\ \bibinfo {author} {\bibfnamefont {Z.-C.}\ \bibnamefont {Gu}},\ }\href
  {https://link.aps.org/doi/10.1103/PhysRevB.95.195147} {\bibfield  {journal}
  {\bibinfo  {journal} {PRB}\ }\textbf {\bibinfo {volume} {95}},\ \bibinfo
  {pages} {195147} (\bibinfo {year} {2017})}\BibitemShut {NoStop}%
\bibitem [{\citenamefont {Cheng}\ \emph
  {et~al.}(2018{\natexlab{b}})\citenamefont {Cheng}, \citenamefont
  {Tantivasadakarn},\ and\ \citenamefont {Wang}}]{Cheng2018}%
  \BibitemOpen
  \bibfield  {author} {\bibinfo {author} {\bibfnamefont {M.}~\bibnamefont
  {Cheng}}, \bibinfo {author} {\bibfnamefont {N.}~\bibnamefont
  {Tantivasadakarn}}, \ and\ \bibinfo {author} {\bibfnamefont {C.}~\bibnamefont
  {Wang}},\ }\href {https://link.aps.org/doi/10.1103/PhysRevX.8.011054}
  {\bibfield  {journal} {\bibinfo  {journal} {PRX}\ }\textbf {\bibinfo {volume}
  {8}},\ \bibinfo {pages} {011054} (\bibinfo {year}
  {2018}{\natexlab{b}})}\BibitemShut {NoStop}%
\bibitem [{\citenamefont {Ando}\ and\ \citenamefont {Fu}(2015)}]{Ando2015}%
  \BibitemOpen
  \bibfield  {author} {\bibinfo {author} {\bibfnamefont {Y.}~\bibnamefont
  {Ando}}\ and\ \bibinfo {author} {\bibfnamefont {L.}~\bibnamefont {Fu}},\
  }\bibfield  {booktitle} {\emph {\bibinfo {booktitle} {Annual Review of
  Condensed Matter Physics}},\ }\href {\doibase
  10.1146/annurev-conmatphys-031214-014501} {\bibfield  {journal} {\bibinfo
  {journal} {Annu. Rev. Condens. Matter Phys.}\ }\textbf {\bibinfo {volume}
  {6}},\ \bibinfo {pages} {361} (\bibinfo {year} {2015})}\BibitemShut {NoStop}%
\bibitem [{\citenamefont {Thorngren}\ and\ \citenamefont
  {Else}(2018)}]{Thorngren2018}%
  \BibitemOpen
  \bibfield  {author} {\bibinfo {author} {\bibfnamefont {R.}~\bibnamefont
  {Thorngren}}\ and\ \bibinfo {author} {\bibfnamefont {D.~V.}\ \bibnamefont
  {Else}},\ }\href {https://link.aps.org/doi/10.1103/PhysRevX.8.011040}
  {\bibfield  {journal} {\bibinfo  {journal} {Phys. Rev. X}\ }\textbf {\bibinfo
  {volume} {8}},\ \bibinfo {pages} {011040} (\bibinfo {year}
  {2018})}\BibitemShut {NoStop}%
\bibitem [{\citenamefont {Jiang}\ and\ \citenamefont {Ran}(2017)}]{Jiang2017}%
  \BibitemOpen
  \bibfield  {author} {\bibinfo {author} {\bibfnamefont {S.}~\bibnamefont
  {Jiang}}\ and\ \bibinfo {author} {\bibfnamefont {Y.}~\bibnamefont {Ran}},\
  }\href {\doibase 10.1103/PhysRevB.95.125107} {\bibfield  {journal} {\bibinfo
  {journal} {Phys. Rev. B}\ }\textbf {\bibinfo {volume} {95}},\ \bibinfo
  {pages} {125107} (\bibinfo {year} {2017})}\BibitemShut {NoStop}%
\bibitem [{\citenamefont {Song}\ \emph
  {et~al.}(2017{\natexlab{a}})\citenamefont {Song}, \citenamefont {Huang},
  \citenamefont {Fu},\ and\ \citenamefont {Hermele}}]{Song2017a}%
  \BibitemOpen
  \bibfield  {author} {\bibinfo {author} {\bibfnamefont {H.}~\bibnamefont
  {Song}}, \bibinfo {author} {\bibfnamefont {S.-J.}\ \bibnamefont {Huang}},
  \bibinfo {author} {\bibfnamefont {L.}~\bibnamefont {Fu}}, \ and\ \bibinfo
  {author} {\bibfnamefont {M.}~\bibnamefont {Hermele}},\ }\href
  {https://link.aps.org/doi/10.1103/PhysRevX.7.011020} {\bibfield  {journal}
  {\bibinfo  {journal} {Phys. Rev. X}\ }\textbf {\bibinfo {volume} {7}},\
  \bibinfo {pages} {011020} (\bibinfo {year} {2017}{\natexlab{a}})}\BibitemShut
  {NoStop}%
\bibitem [{\citenamefont {Huang}\ \emph {et~al.}(2017)\citenamefont {Huang},
  \citenamefont {Song}, \citenamefont {Huang},\ and\ \citenamefont
  {Hermele}}]{Huang2017b}%
  \BibitemOpen
  \bibfield  {author} {\bibinfo {author} {\bibfnamefont {S.-J.}\ \bibnamefont
  {Huang}}, \bibinfo {author} {\bibfnamefont {H.}~\bibnamefont {Song}},
  \bibinfo {author} {\bibfnamefont {Y.-P.}\ \bibnamefont {Huang}}, \ and\
  \bibinfo {author} {\bibfnamefont {M.}~\bibnamefont {Hermele}},\ }\href
  {https://link.aps.org/doi/10.1103/PhysRevB.96.205106} {\bibfield  {journal}
  {\bibinfo  {journal} {Phys. Rev. B}\ }\textbf {\bibinfo {volume} {96}},\
  \bibinfo {pages} {205106} (\bibinfo {year} {2017})}\BibitemShut {NoStop}%
\bibitem [{\citenamefont {Lu}\ \emph {et~al.}(2017)\citenamefont {Lu},
  \citenamefont {Shi},\ and\ \citenamefont {Lu}}]{Lu2017c}%
  \BibitemOpen
  \bibfield  {author} {\bibinfo {author} {\bibfnamefont {F.}~\bibnamefont
  {Lu}}, \bibinfo {author} {\bibfnamefont {B.}~\bibnamefont {Shi}}, \ and\
  \bibinfo {author} {\bibfnamefont {Y.-M.}\ \bibnamefont {Lu}},\ }\href
  {http://stacks.iop.org/1367-2630/19/i=7/a=073002} {\bibfield  {journal}
  {\bibinfo  {journal} {New Journal of Physics}\ }\textbf {\bibinfo {volume}
  {19}},\ \bibinfo {pages} {073002} (\bibinfo {year} {2017})}\BibitemShut
  {NoStop}%
\bibitem [{\citenamefont {{Rasmussen}}\ and\ \citenamefont
  {{Lu}}(2018)}]{Rasmussen2018}%
  \BibitemOpen
  \bibfield  {author} {\bibinfo {author} {\bibfnamefont {A.}~\bibnamefont
  {{Rasmussen}}}\ and\ \bibinfo {author} {\bibfnamefont {Y.-M.}\ \bibnamefont
  {{Lu}}},\ }\href@noop {} {\bibfield  {journal} {\bibinfo  {journal} {ArXiv
  e-prints}\ } (\bibinfo {year} {2018})},\ \Eprint
  {http://arxiv.org/abs/1809.07325} {arXiv:1809.07325 [cond-mat.str-el]}
  \BibitemShut {NoStop}%
\bibitem [{\citenamefont {{Shiozaki}}\ \emph {et~al.}(2018)\citenamefont
  {{Shiozaki}}, \citenamefont {{Zhaoxi Xiong}},\ and\ \citenamefont
  {{Gomi}}}]{Shiozaki2018}%
  \BibitemOpen
  \bibfield  {author} {\bibinfo {author} {\bibfnamefont {K.}~\bibnamefont
  {{Shiozaki}}}, \bibinfo {author} {\bibfnamefont {C.}~\bibnamefont {{Zhaoxi
  Xiong}}}, \ and\ \bibinfo {author} {\bibfnamefont {K.}~\bibnamefont
  {{Gomi}}},\ }\href@noop {} {\bibfield  {journal} {\bibinfo  {journal} {ArXiv
  e-prints}\ } (\bibinfo {year} {2018})},\ \Eprint
  {http://arxiv.org/abs/1810.00801} {arXiv:1810.00801 [cond-mat.str-el]}
  \BibitemShut {NoStop}%
\bibitem [{\citenamefont {{Else}}\ and\ \citenamefont
  {{Thorngren}}(2018)}]{Else2018}%
  \BibitemOpen
  \bibfield  {author} {\bibinfo {author} {\bibfnamefont {D.~V.}\ \bibnamefont
  {{Else}}}\ and\ \bibinfo {author} {\bibfnamefont {R.}~\bibnamefont
  {{Thorngren}}},\ }\href@noop {} {\bibfield  {journal} {\bibinfo  {journal}
  {ArXiv e-prints}\ } (\bibinfo {year} {2018})},\ \Eprint
  {http://arxiv.org/abs/1810.10539} {arXiv:1810.10539 [cond-mat.str-el]}
  \BibitemShut {NoStop}%
\bibitem [{\citenamefont {{Song}}\ \emph {et~al.}(2018)\citenamefont {{Song}},
  \citenamefont {{Fang}},\ and\ \citenamefont {{Qi}}}]{Song2018a}%
  \BibitemOpen
  \bibfield  {author} {\bibinfo {author} {\bibfnamefont {Z.}~\bibnamefont
  {{Song}}}, \bibinfo {author} {\bibfnamefont {C.}~\bibnamefont {{Fang}}}, \
  and\ \bibinfo {author} {\bibfnamefont {Y.}~\bibnamefont {{Qi}}},\ }\href@noop
  {} {\bibfield  {journal} {\bibinfo  {journal} {ArXiv e-prints}\ } (\bibinfo
  {year} {2018})},\ \Eprint {http://arxiv.org/abs/1810.11013} {arXiv:1810.11013
  [cond-mat.str-el]} \BibitemShut {NoStop}%
\bibitem [{\citenamefont {Parameswaran}\ and\ \citenamefont
  {Wan}(2017)}]{Parameswaran2017}%
  \BibitemOpen
  \bibfield  {author} {\bibinfo {author} {\bibfnamefont {S.~A.}\ \bibnamefont
  {Parameswaran}}\ and\ \bibinfo {author} {\bibfnamefont {Y.}~\bibnamefont
  {Wan}},\ }\href@noop {} {\bibfield  {journal} {\bibinfo  {journal} {Physics}\
  }\textbf {\bibinfo {volume} {10}},\ \bibinfo {pages} {132} (\bibinfo {year}
  {2017})}\BibitemShut {NoStop}%
\bibitem [{\citenamefont {Benalcazar}\ \emph
  {et~al.}(2017{\natexlab{a}})\citenamefont {Benalcazar}, \citenamefont
  {Bernevig},\ and\ \citenamefont {Hughes}}]{Benalcazar2017}%
  \BibitemOpen
  \bibfield  {author} {\bibinfo {author} {\bibfnamefont {W.~A.}\ \bibnamefont
  {Benalcazar}}, \bibinfo {author} {\bibfnamefont {B.~A.}\ \bibnamefont
  {Bernevig}}, \ and\ \bibinfo {author} {\bibfnamefont {T.~L.}\ \bibnamefont
  {Hughes}},\ }\href
  {http://science.sciencemag.org/content/357/6346/61.abstract} {\bibfield
  {journal} {\bibinfo  {journal} {Science}\ }\textbf {\bibinfo {volume}
  {357}},\ \bibinfo {pages} {61} (\bibinfo {year}
  {2017}{\natexlab{a}})}\BibitemShut {NoStop}%
\bibitem [{\citenamefont {Benalcazar}\ \emph
  {et~al.}(2017{\natexlab{b}})\citenamefont {Benalcazar}, \citenamefont
  {Bernevig},\ and\ \citenamefont {Hughes}}]{Benalcazar2017a}%
  \BibitemOpen
  \bibfield  {author} {\bibinfo {author} {\bibfnamefont {W.~A.}\ \bibnamefont
  {Benalcazar}}, \bibinfo {author} {\bibfnamefont {B.~A.}\ \bibnamefont
  {Bernevig}}, \ and\ \bibinfo {author} {\bibfnamefont {T.~L.}\ \bibnamefont
  {Hughes}},\ }\href {\doibase 10.1103/PhysRevB.96.245115} {\bibfield
  {journal} {\bibinfo  {journal} {Phys. Rev. B}\ }\textbf {\bibinfo {volume}
  {96}},\ \bibinfo {pages} {245115} (\bibinfo {year}
  {2017}{\natexlab{b}})}\BibitemShut {NoStop}%
\bibitem [{\citenamefont {Song}\ \emph
  {et~al.}(2017{\natexlab{b}})\citenamefont {Song}, \citenamefont {Fang},\ and\
  \citenamefont {Fang}}]{Song2017}%
  \BibitemOpen
  \bibfield  {author} {\bibinfo {author} {\bibfnamefont {Z.}~\bibnamefont
  {Song}}, \bibinfo {author} {\bibfnamefont {Z.}~\bibnamefont {Fang}}, \ and\
  \bibinfo {author} {\bibfnamefont {C.}~\bibnamefont {Fang}},\ }\href {\doibase
  10.1103/PhysRevLett.119.246402} {\bibfield  {journal} {\bibinfo  {journal}
  {Phys. Rev. Lett.}\ }\textbf {\bibinfo {volume} {119}},\ \bibinfo {pages}
  {246402} (\bibinfo {year} {2017}{\natexlab{b}})}\BibitemShut {NoStop}%
\bibitem [{\citenamefont {Langbehn}\ \emph {et~al.}(2017)\citenamefont
  {Langbehn}, \citenamefont {Peng}, \citenamefont {Trifunovic}, \citenamefont
  {von Oppen},\ and\ \citenamefont {Brouwer}}]{Langbehn2017}%
  \BibitemOpen
  \bibfield  {author} {\bibinfo {author} {\bibfnamefont {J.}~\bibnamefont
  {Langbehn}}, \bibinfo {author} {\bibfnamefont {Y.}~\bibnamefont {Peng}},
  \bibinfo {author} {\bibfnamefont {L.}~\bibnamefont {Trifunovic}}, \bibinfo
  {author} {\bibfnamefont {F.}~\bibnamefont {von Oppen}}, \ and\ \bibinfo
  {author} {\bibfnamefont {P.~W.}\ \bibnamefont {Brouwer}},\ }\href {\doibase
  10.1103/PhysRevLett.119.246401} {\bibfield  {journal} {\bibinfo  {journal}
  {Phys. Rev. Lett.}\ }\textbf {\bibinfo {volume} {119}},\ \bibinfo {pages}
  {246401} (\bibinfo {year} {2017})}\BibitemShut {NoStop}%
\bibitem [{\citenamefont {Chen}\ \emph {et~al.}(2014)\citenamefont {Chen},
  \citenamefont {Lu},\ and\ \citenamefont {Vishwanath}}]{Chen2014}%
  \BibitemOpen
  \bibfield  {author} {\bibinfo {author} {\bibfnamefont {X.}~\bibnamefont
  {Chen}}, \bibinfo {author} {\bibfnamefont {Y.-M.}\ \bibnamefont {Lu}}, \ and\
  \bibinfo {author} {\bibfnamefont {A.}~\bibnamefont {Vishwanath}},\ }\href
  {http://dx.doi.org/10.1038/ncomms4507} {\bibfield  {journal} {\bibinfo
  {journal} {Nat Commun}\ }\textbf {\bibinfo {volume} {5}},\  (\bibinfo {year}
  {2014})}\BibitemShut {NoStop}%
\bibitem [{\citenamefont {Kane}\ and\ \citenamefont {Mele}(2005)}]{Kane2005}%
  \BibitemOpen
  \bibfield  {author} {\bibinfo {author} {\bibfnamefont {C.~L.}\ \bibnamefont
  {Kane}}\ and\ \bibinfo {author} {\bibfnamefont {E.~J.}\ \bibnamefont
  {Mele}},\ }\href {http://link.aps.org/doi/10.1103/PhysRevLett.95.226801}
  {\bibfield  {journal} {\bibinfo  {journal} {Phys. Rev. Lett.}\ }\textbf
  {\bibinfo {volume} {95}},\ \bibinfo {pages} {226801} (\bibinfo {year}
  {2005})}\BibitemShut {NoStop}%
\bibitem [{\citenamefont {Qi}(2013)}]{Qi2013}%
  \BibitemOpen
  \bibfield  {author} {\bibinfo {author} {\bibfnamefont {X.-L.}\ \bibnamefont
  {Qi}},\ }\href {http://stacks.iop.org/1367-2630/15/i=6/a=065002} {\bibfield
  {journal} {\bibinfo  {journal} {New Journal of Physics}\ }\textbf {\bibinfo
  {volume} {15}},\ \bibinfo {pages} {065002} (\bibinfo {year}
  {2013})}\BibitemShut {NoStop}%
\bibitem [{\citenamefont {Ryu}\ and\ \citenamefont {Zhang}(2012)}]{Ryu2012}%
  \BibitemOpen
  \bibfield  {author} {\bibinfo {author} {\bibfnamefont {S.}~\bibnamefont
  {Ryu}}\ and\ \bibinfo {author} {\bibfnamefont {S.-C.}\ \bibnamefont
  {Zhang}},\ }\href {http://link.aps.org/doi/10.1103/PhysRevB.85.245132}
  {\bibfield  {journal} {\bibinfo  {journal} {Phys. Rev. B}\ }\textbf {\bibinfo
  {volume} {85}},\ \bibinfo {pages} {245132} (\bibinfo {year}
  {2012})}\BibitemShut {NoStop}%
\bibitem [{\citenamefont {Yao}\ and\ \citenamefont {Ryu}(2013)}]{Yao2013}%
  \BibitemOpen
  \bibfield  {author} {\bibinfo {author} {\bibfnamefont {H.}~\bibnamefont
  {Yao}}\ and\ \bibinfo {author} {\bibfnamefont {S.}~\bibnamefont {Ryu}},\
  }\href {http://link.aps.org/doi/10.1103/PhysRevB.88.064507} {\bibfield
  {journal} {\bibinfo  {journal} {Phys. Rev. B}\ }\textbf {\bibinfo {volume}
  {88}},\ \bibinfo {pages} {064507} (\bibinfo {year} {2013})}\BibitemShut
  {NoStop}%
\bibitem [{\citenamefont {Kitaev}(2001)}]{Kitaev2001}%
  \BibitemOpen
  \bibfield  {author} {\bibinfo {author} {\bibfnamefont {A.~Y.}\ \bibnamefont
  {Kitaev}},\ }\href {http://stacks.iop.org/1063-7869/44/i=10S/a=S29}
  {\bibfield  {journal} {\bibinfo  {journal} {Physics-Uspekhi}\ }\textbf
  {\bibinfo {volume} {44}},\ \bibinfo {pages} {131} (\bibinfo {year}
  {2001})}\BibitemShut {NoStop}%
\bibitem [{\citenamefont {Fidkowski}\ and\ \citenamefont
  {Kitaev}(2010)}]{Fidkowski2010}%
  \BibitemOpen
  \bibfield  {author} {\bibinfo {author} {\bibfnamefont {L.}~\bibnamefont
  {Fidkowski}}\ and\ \bibinfo {author} {\bibfnamefont {A.}~\bibnamefont
  {Kitaev}},\ }\href {http://link.aps.org/doi/10.1103/PhysRevB.81.134509}
  {\bibfield  {journal} {\bibinfo  {journal} {Phys. Rev. B}\ }\textbf {\bibinfo
  {volume} {81}},\ \bibinfo {pages} {134509} (\bibinfo {year}
  {2010})}\BibitemShut {NoStop}%
\end{thebibliography}%

\end{document}